%% file: Banerjee_Karfa_Intro_Array_Program_Verification.tex
\documentclass[conference]{IEEEtran}
\IEEEoverridecommandlockouts
\usepackage{cite}
\usepackage{amsmath,amssymb,amsfonts}
\usepackage{algorithmic}
\usepackage{graphicx}
\usepackage{textcomp}
\usepackage{xcolor}
\def\BibTeX{{\rm B\kern-.05em{\sc i\kern-.025em b}\kern-.08em
    T\kern-.1667em\lower.7ex\hbox{E}\kern-.125emX}}
\begin{document}

\title{A Quick Introduction to Functional Verification of Array-Intensive Programs}

\author{\IEEEauthorblockN{Kunal Banerjee}
\IEEEauthorblockA{\textit{Parallel Computing Lab} \\
\textit{Intel Corporation}\\
Bangalore, India \\
kunal.banerjee@intel.com}
\and
\IEEEauthorblockN{Chandan Karfa}
\IEEEauthorblockA{\textit{Dept. of Computer Sc. \& Engg.} \\
\textit{IIT Guwahati}\\
Guwahati, India \\
ckarfa@iitg.ac.in}
}

\maketitle

\begin{abstract}
Array-intensive programs are often amenable to parallelization across many cores
on a single machine as well as scaling across multiple machines and hence are
well explored, especially in the domain of high-performance computing.
These programs typically undergo loop transformations and arithmetic 
transformations in addition to parallelizing transformations.
Although a lot of effort has been invested in improving parallelizing compilers,
experienced programmers still resort to hand-optimized transformations which
is typically followed by careful tuning of the transformed program to finally
obtain the optimized program.
Therefore, it is critical to verify that the functional correctness of an original 
sequential program is not sacrificed during the process of optimization.
In this paper, we cover important literature on functional verification of array-intensive 
programs which we believe can be a good starting point for one interested in this field.
\end{abstract}

\begin{IEEEkeywords}
formal verification, functional correctness, array-intensive programs, optimization
\end{IEEEkeywords}

\section{Introduction}\label{Sec:intro}

Recent days have seen a boost in high-performance computing due to the introduction of
high performant GPUs and specially designed coprocessors, such as Xeon Phi.
The primary workloads which have benefited from these new hardware are the array-intensive
programs (also known as dataflow programs).
These programs are amenable to parallelization and various application programming interfaces
(APIs) exist which help in exploiting the many cores available nowadays on a machine, such as
OpenMP~\cite{openmp} and OpenACC~\cite{openacc}, and also to parallelize across multiple
machines, such as MPI~\cite{mpi} and OpenCL~\cite{opencl}.
However, although a lot of research has gone into improving parallelizing
compilers~\cite{polaris,pluto}, these compilers are still not popular among
experienced programmers who prefer applying hand-crafted transformations to a
sequential program, followed by a tuning phase of the transformed program which
finally results in an optimally performing program. 
In such a case, it is even more crucial to verify if the parallelizing transformation
along with other optimizations, like loop transformations and arithmetic transformations,
preserve the semantic equivalence of the sequential code.
In this paper, we present a short survey of the significant work done in the field
of functional verification of array-intensive programs.

Additionally, it is important to note that customized hardware has received a lot of attention
recently due to the rapid proliferation of deep learning -- be it coarse-grained reconfigurable
architecture (CGRA)~\cite{cgraNN1,cgraNN2}, FPGA~\cite{fpgaNN1} or systolic array~\cite{systolicNN1}.
A lot of compiler frameworks have also come up to provide support for the heterogeneous architecture
that may be used to train the neural networks, for example, Google's Tensorflow XLA~\cite{xla},
Amazon endorsed NNVM~\cite{nnvm} and Facebook's Glow~\cite{glow}.
New programming methodologies are also being proposed to further boost programmer productivity by
exploiting underlying architecture with minimal programming overhead; the work reported in~\cite{T2S},
for instance, introduces an abstraction layer in the program called T2S (temporal to spatial) which
takes a temporal definition (basically, the function to be computed) and generates its spatial mapping
(i.e., decompose the specified function and map the decomposed pieces onto a spatial architecture, e.g., 
CGRA or FPGA).
There is always a possibility that the compiler, e.g.~\cite{xla,nnvm,glow}, may have an implementation bug 
or the specification provided, say in T2S~\cite{T2S}, has some inherent logical mistake because of which
the prescribed functionality cannot be efficiently mapped to a given FPGA.
We found that literature on verification of specifically tailored transformations for mapping neural networks
onto heterogeneous architecture (which primarily consists of the transformations discussed in this paper) is
still at a nascent stage.
In particular, the work reported in~\cite{dlvm} offloads the onus of compiler transformation verification
by translating its internal representation (IR) to the well-known LLVM~\cite{llvm} IR and then leveraging
existing verification techniques for LLVM compiler.
Therefore, we believe that our paper should also appeal to those who are interested in designing and/or
verifying compiler transformations directed towards deep learning applications or artificial intelligence, 
in general.

The rest of the paper is organized as follows. 
Section~\ref{sec:optimize} illustrates the benefits of applying loop transformations and
arithmetic transformations to array-intensive programs. 
Section~\ref{sec:method} covers the various significant work done for verifying such
programs with the help of an example.
Section~\ref{sec:concl} concludes the paper.

\section{Optimizing transformations}\label{sec:optimize}

\subsection{Applications of loop transformations}\label{sec:loop}

Multimedia and signal processing applications have witnessed extensive application
of loop transformations and arithmetic transformations.
In the following, we study several applications of loop transformations techniques
during (embedded) software design.

The effects of loop transformations on system power has been studied extensively.
The work reported in~\cite{337425} underlines the effect of loop fusion, loop
fission, loop unrolling, loop tiling, and scalar expansion on energy consumption.
The futility of applying conventional data locality oriented code transformations
for minimizing disk power consumption has been showcased in~\cite{1077655}.
As a remedy, the authors of~\cite{1077655} suggest how both code restructuring and
data locality optimization should be taken into consideration for designing a disk
layout aware application optimization strategy.
Specifically, the authors focus on three optimizations -- loop fusion/fission,
loop tiling and linear optimizations -- for code restructuring and advocate a
unified optimizer that targets disk power management by applying these transformations.
In~\cite{1142163}, the authors focus on an MPSoC architecture with a banked memory
system.
For this architecture, they demonstrate how code and data optimizations assist in
reducing memory energy consumption for embedded applications with regular data
access patterns.
The work in~\cite{Xue:2007:MCP:1278480.1278536} achieves minimization of the
data memory requirements of processors by using a memory-conscious loop
parallelization strategy.
A data space-oriented tiling (DST) approach is proposed in~\cite{kadayifK05tecs}
whereby the data space is logically divided into chunks called data tiles.
DST has the potentiality to achieve better results than conventional loop tiling
because it exploits inter-nest data locality since the data space is common across
all loop nests that access it.
A global approach to tackle data locality problem is prescribed in~\cite{li05workload} 
which evaluates all the loop nests in an application to be run in an embedded MPSoC
simultaneously and schedules the different constituent modules accordingly for
parallel execution.
In the context of an embedded chip multiprocessor, the method described
in~\cite{1118342} underlines how reliability against transient errors can be
enhanced without sacrificing execution time by replicating some of the operations
being executed on active processors onto (otherwise) idle processors.

Loop transformations have found application in the design of system memory as well.
For example, the authors of~\cite{bouchebaba07} explain a method in the context of
multimedia applications running on MPSoCs that can reduce cache misses and also
cache size.
Specifically, loop fusion and loop tiling are employed to minimize cache misses,
whereas a novel buffer allocation strategy is used to reduce cache size.
This work is extended in~\cite{1241831} to handle dependence-free arrays additionally.
Here an input-conscious tiling scheme for off-chip memory access optimization
is proposed.
The authors showcase that the input arrays play an as important role as the arrays
with data dependencies when the objective is memory access optimization instead of
parallelism extraction.
Data reuse is a key process that may potentially reduce external memory access
by exploiting the memory hierarchy.
Loop transformations for data locality and memory hierarchy allocation are important
procedures in the optimization flow for data reuse.
A global approach that provides optimal results on external memory bandwidth and
on-chip data reuse buffer size by combining loop transformations and memory
hierarchy allocation can be found in~\cite{memoryAlloc}.
An extension of this work is presented in~\cite{memoryAllocation} that optimizes
on-chip memory allocation by loop transformations in the imperfectly nested loops.
A dynamic loop tiling strategy is proposed in~\cite{sanket11,sanket13} to enhance
cache locality and obtain coarse-grained parallelism.

In~\cite{1053675}, the authors undertake the challenge of reducing the total energy
while maintaining the performance requirements for application with
multi-dimensional nested loops. 
They have demonstrated that an adaptive loop parallelization strategy along with
idle processor shut down and pre-activation can be critical in minimizing energy 
consumption without increasing execution time.
The objective of the paper~\cite{qiu08} is also the same as that of~\cite{1053675}. 
However, they apply loop fusion and multi-functional unit scheduling techniques to achieve that.

Loops containing nested conditional blocks can pose serious challenge to
compilers while producing optimized code.
This problem is tackled in~\cite{springerlink:10.1007}.
This work statically analyzes the Boolean conditions appearing at branching states
in the control flow of a program using a novel interval analysis technique.
The outputs of interval analysis integrated with those of loop dependency are used to
segregate the iteration space of the nested loops.

A survey on application of loop transformations in data and memory optimization in embedded
system can be found in~\cite{Panda:2001:DMO:375977.375978},
whereas the benefits of these transformations outlined in~\cite{baconSharp} are targeted
for a more general software design. 
For details on some of the pioneering work on program transformations targeting
reduction of energy consumption in dataflow programs, one may refer 
to~\cite{imecbk,palkovic09}.
Another work~\cite{FrabouletKM01} tries to reduce the use of temporary arrays,
which may eventually result in better register usage, by using loop fusion technique
in multimedia applications before hardware/software partitioning is carried out.
Loop transformations have also been applied to improve performance in coarse-grained
reconfigurable architecture~\cite{loopCgra}.
Applications of loop transformations to parallelize sequential code targeting
embedded multi-core systems are given in~\cite{7084524,loopMultiCore}.
Interested readers are encouraged to look into
\cite{Simunic:2000:SCO:501790.501831,1026005,Brandolese:2004:AME:968880.969253,Ghodrat:2008:CFO:1450095.1450120,falkbook} 
for several other loop transformation techniques and their effects on system design.

\subsection{Applications of arithmetic transformations}\label{sec:arith}

Compiler optimizations often involve several arithmetic transformations based on algebraic properties of the
operator such as associativity, commutativity and distributivity, arithmetic expression simplification,
constant folding, common sub-expression elimination, renaming, dead code elimination, copy propagation and
operator strength reduction, etc.
For example, the work~\cite{potkonjakRetiming} demonstrates how applying retiming, algebraic and redundancy manipulation transformations can drastically improve 
the performance of embedded systems.
The authors of~\cite{zoryPACT} investigate source-to-source algebraic
transformations which aid in decreasing the execution time of expression evaluation;
its benefit on performance has been recorded on many computationally intensive
applications.
They, basically, propose two algorithms based on factorization and multiply-add extraction heuristics to replace traditional associative commutative pattern-matching techniques. 
Another method that minimizes operation cost based on loop-invariant code motion and
operator strength reduction is reported in~\cite{sguptaDATE00}.
An in-depth experimental analysis on the effectiveness of such source-level
transformations at the level of number of execution cycles, before and after
applying the optimizations, is given in~\cite{sguptaDATE00} for two real-life
multimedia application kernels.
Application of algebraic transformations to minimize critical path length in the
domain of computationally intensive applications is proposed in~\cite{landwehr97}.
Apart from standard algebraic transformations such as commutativity, associativity
and distributivity, they also introduce two hardware related transformations based
on operator strength reduction and constant unfolding.
A set of transformations such as common sub-expression elimination, renaming, dead
code elimination and copy propagation are applied along with code motion
transformations in the pre-synthesis and scheduling phase of high-level synthesis in
the SPARK tool~\cite{spark03,sGuptaACM}.
The potential of arithmetic transformations on FPGAs is studied in~\cite{eoapndg03}.
It has been shown that operator strength reduction and storage reuse reduce the area
of the circuit and hence the power consumption in FPGA.
The transformations like height reduction and variable renaming reduce the total
number of clock cycles required to execute the programs in FPGAs, whereas expression
splitting and resource sharing reduce the clock period of the circuits.

\section{Verification of optimizing transformations}\label{sec:method}

\begin{figure*}[th]
\begin{minipage}[b]{6cm}
{\small
\begin{verbatim}
void program1(
 int A[], int B[], int C[]) {
  int i, I1[500], I2[500];
  S0: C[0] = A[0] + 2;
  S1: C[0] += B[0] + 2;
  for (int i=1; i<N; i++) {     
    S2: I1[i] = A[i] * C[i-1];
    S3: I2[i] = B[i] * C[i-1];
    S4: C[i] = I1[i] + I2[i];
} }

           (a)
\end{verbatim}}
\end{minipage}
\hfill
\begin{minipage}[b]{6cm}
{\small
\begin{verbatim}
void program2(
 int A[], int B[], int C[]){
  int i, I[500];
  T0: C[0] = A[0] + B[0] + 4;
#pragma omp parallel for
  for (i=1; i<N; i++) {
    T1: I[i] = A[i] + B[i];
  }
#pragma omp parallel for
  for (i=1; i<N; i++) {
    T2: C[i] = I[i] * C[i-1];
} }
              (b)
\end{verbatim}}
\end{minipage}
\caption{(a) Original sequential program. (b) Transformed parallel program.}
\label{Fig:example}
\end{figure*}

Let us consider the two functions shown in Figure~\ref{Fig:example}.
The function shown in Figure~\ref{Fig:example}(b) has been obtained from that
of Figure~\ref{Fig:example}(a) by applying the following transformations.\\
\textit{Application of loop transformation:} The \texttt{for} loop in \texttt{program1} has been 
split into two (an instance of loop fission).\\
\textit{Application of arithmetic transformations:} The statements \texttt{S0}
and \texttt{S1} in \texttt{program1} have been merged into one statement
\texttt{T0} in \texttt{program2} -- it is regarded as a linear arithmetic transformation. 
Moreover, distributive property of multiplication over addition has been applied
to reduce the number of temporary array variables from two (\texttt{I1} and
\texttt{I2}) to one (\texttt{I}) -- it is regarded as a non-linear arithmetic
transformation because it involves multiplication of two array variables.\\
\textit{Application of parallelizing transformations:} OpenMP directives have been introduced in \texttt{program2}
to make the execution of its \texttt{for} loops parallel.\\
Also note that the functions contain recurrence because the array \texttt{C} is defined
in terms of previously defined elements of the same array.
Now let us go through the various methods reported in literature for verifying
these individual transformations or some combination of these.

Verification of loop transformations on array-intensive programs is a well studied problem.
Some of these target transformation specific verification rules.
The techniques reported in~\cite{voc,pnueli05}, for example, proposed permutation
rules for verifying loop tiling, loop reversal, loop skewing, loop interchange
transformations in their translation validation approach.
The rule set is further enhanced in~\cite{loopEntcs,tvoc}.
The primary issue with this approach is that the method had to rely on the hint provided by the compiler.
The verifier needs the list of transformations that have been applied and the order in which they have been
applied from the synthesis tool.
Also, completeness of the verifier depends on the completeness of the rule set and
therefore enhancement of the repository of transformations necessitates enhancement
of the rule set.

The concept of fractal symbolic analysis is introduced in~\cite{fractal}.
The idea is to reduce the gap between the source and the transformed programs by
applying simplification rules repeatedly until the two programs become similar
enough to allow a proof by symbolic analysis.
The rules are similar to the ones proposed by~\cite{voc,pnueli05}.
This method combines some of the power of symbolic analysis with the tractability of dependence analysis.
The applicability of this technique again depends on the robustness of the provided
rule set.

The design of a fully automatic verifier for loop transformations can be found
in~\cite{samsom}.
Preservation of data dependencies between the original and the transformed
programs at a statement-level forms the central idea in this work.
This method, however, does not have provision to handle arithmetic transformations.
Since it is common that the arithmetic transformations and the loop transformations
are applied together, direct correspondence between the
statement classes of the original and the transformed programs does not always hold
as necessitated by~\cite{samsom}.

Off-the-shelf SMT solvers, such as CVC4~\cite{cvc4}, Yices~\cite{yices}, or theorem
provers, such as ACL2~\cite{acl2}, have also been demonstrated to verify loop
transformations and arithmetic transformations~\cite{icare13}.
It is more or less straightforward to model the equivalence of two programs
with a formula. The validity of the formula can be checked by a SMT
solver or theorem prover~\cite{icare13}; if the formula is found to be valid then
the two programs are indeed equivalent.
It is to be noted that although the SMT solvers and the theorem prover can be 
effective in handling linear arithmetic, presence of non-linear arithmetic often
makes these tools falter in proving the equivalence; in such scenarios, these tools
either output ``unknown'' indicating that they failed to prove either equivalence
or non-equivalence of the programs or they time out without producing an output. 
It has been shown in\cite{icare13} that state-of-the-art SMT solvers failed to verify 
most of the loop transformations and arithmetic transformations. 

The method developed in~\cite{kcs02,kcscc05,shashidhar05,shashidhar08} assesses a
restricted class  of programs with affine indices and
bounds, static control flow, single assignment form and valid schedule.      
This method presents a translation validation algorithm for verifying loop
transformations, where the source and the transformed programs are modeled as
Array Data Dependence Graphs (ADDGs).
This method is promising since it is capable of handling most loop transformations
without requiring any additional information from the compilers (or human expert).
The primary limitations of this ADDG based verification technique are its inability 
to handle recurrences and arithmetic transformations.
The ADDG based method is extended in two directions initially.  
In one direction, Verdoolaege et al. in~\cite{wideningCAV09,widening12} enhanced 
the method to handle recurrences in programs.
In another direction, Karfa et al.~\cite{isvlsi11,karfaTcad13} enhanced it to handle 
arithmetic transformations.

Specifically, in~\cite{wideningCAV09,widening12}, the authors have modelled the programs 
as dependence graphs (DGs). 
Before delving into its details, note that a recurrence involves \textit{induction case(s)}, 
e.g. statement \texttt{T2} in Figure~\ref{Fig:example}(b), whereby the members of an
array is defined in terms of previously defined members of itself, and \textit{basis case(s)},
e.g. statement \texttt{T1} in Figure~\ref{Fig:example}(b).
Typically, proving the basis cases in the DGs obtained from the original and the transformed
programs is straightforward involving symbolic substitution of the temporary arrays by input
arrays.
However, proving equivalence between the induction cases can be convoluted and hence this
verification procedure takes an optimistic approach and initially considers that the induction
cases in the two DGs are possibly equivalent and proceeds in a forward pass (using an operation
called \textit{widening});
the proof obligations that remain pending during the forward pass are tried to be resolved during
a backward pass (using an inverse operation called \textit{narrowing}).
This two pass approach adopted in~\cite{wideningCAV09,widening12} is found to be effective
to handle a wide range of programs containing recurrences.
In fact, this method has been proved to be successful in verying realistic multimedia systems in~\cite{widening_J10}.

In other direction, a slice-level equivalence of ADDGs is proposed in~\cite{isvlsi11,karfaTcad13} (as opposed to
path-level equivalence of~\cite{shashidhar05,shashidhar08}), to handle arithmetic transformations such as, constant
unfolding, common sub-expression elimination, distributive transformations,
arithmetic expression simplification, etc., along with loop transformations.
This method additionally incorporates a normalization
technique~\cite{dsNorm} extended suitably to represent data transformations.
It has also been adopted in checking correctness of transformations on the 
Kahn process networks (KPNs) for multimedia and signal processing applications~\cite{karfaVlsid13}.
Handling recurrences, however, remains as the main limitation of this technique. 
Recurrences create cycles in the ADDG representation of a program. In the presence of loops in an ADDG, 
obtaining the closed-form representations of the data dependence is hard.
The work reported in~\cite{lctes16}, for the first time, proposes a unified method
to verify loop and arithmetic transformation as well as recurrence.
In~\cite{lctes16}, the verifier first identifies cyclic subgraphs in the ADDGs
(arising from recurrences), then for each cyclic subgraph in the original ADDG,
it tries to find an equivalent cyclic subgraph in the transformed ADDG; 
if one-to-one correspondence is found for all the cyclic subgraphs in the two ADDGs,
then the entire ADDGs, with the cyclic subgraphs replaced by some uninterpreted functions 
(and thus reduced to cycle-free ADDGs), are compared in an identical
way as mentioned in~\cite{karfaTcad13}.

None of the above methods, however, handle parallelizing transformations and loop vectorization.
An early work~\cite{Flanagan2002} proposes a static checker for analysis of a number of 
thread synchronization issues. 
Another work~\cite{Deng2002} proposes a similar approach for synthesizing
synchronization implementation from a high level specification.
However, the methods described in~\cite{Flanagan2002,Deng2002} do not explicitly
target loop transformations and cannot handle arithmetic transformations at all.
The method of~\cite{lctes16} was extended in~\cite{array16} to handle
data races which may arise on introducing parallelizing transformations. 
In~\cite{array16}, the authors model the programs as coloured-ADDGs.
Specifically, nodes with different colours are maintained to capture the
different parallelized regions of the program and if some thread originating in
some coloured region of the program is found to be able to enter a differently
coloured region of the program (signifying that no synchronization barrier exists between
these two regions), then the method declares that parallelized program as unsafe
because data race exists in that program. In \cite{fase15}, a method is proposed to detect if a loop can be parallelized or not. The authors have used seperation logic to detect that.  In addition, the method also identifies the synchronization points which is required for parallel programs. Dutta et. al ~\cite{dutta16,Dutta17} has extended the work of 
Verdoolaege et. al. in~\cite{wideningCAV09,widening12} for verification of loop vectorization and loop 
parallelization transformations. 
Specifically, they have shown how to construct a DG from the loop vectorized and the loop parallelized programs. 
The proposed work then applies the DG based method proposed in~\cite{wideningCAV09,widening12} to check the equivalence. 
They have also enhanced the method proposed in~\cite{wideningCAV09,widening12} to handle loop collapsing transformation. 

\section{Conclusion}\label{sec:concl}
Table~\ref{Tab:compare} provides a list of the pros and cons of several key
literature mentioned here.
It is worth noting that while supporting as many transformations as possible is a
desired property for a method, needing hints from the compiler to do so is
typically considered to be undesirable because it requires explicit instrumentation
for each compiler (synthesizing tool); moreover, a human expert may not always
document one's intuition for applying a transformation methodically.
Although~~\cite{lctes16} tries to bridge the gap between handling recurrences and arithmetic transformations, 
the treatment of recurrence in~\cite{wideningCAV09,widening12} is more robust, for example, the method 
of~\cite{widening12} can additionally handle cases of mutual recurrence (also known as, co-induction). 
The arithmetic transformations are elegantly handled in~\cite{isvlsi11,karfaTcad13}.
Therefore, it would be an interesting future work to apply the normalization techniques proposed 
in~\cite{isvlsi11,karfaTcad13} on the DG based method proposed in~\cite{wideningCAV09,widening12} to handle both 
recurrence and arithmetic transformations elegantly. 
In addition, the work presented in~\cite{dutta16,Dutta17} can be used prior to this method to handle 
loop vectorization and loop parallelization transformations.
We further envision that the verification techniques discussed here are also
effective in verifying transformations applied by compilers targeting deep
learning applications.

\begin{table*}[t]
\caption{A comparison among different methods based on transformations supported}
\label{Tab:compare}
\centering
\begin{tabular}{| c | c | c | c | c | c | c |}
\hline
Lit                    & Need hint    & Loop         & Recur        & Arith        & Vector   & Parallel \\
\hline
\cite{tvoc}            & $\checkmark$ & $\checkmark$ & $\times$     & $\times$     & $\times$ & $\times$ \\
\cite{fractal}         & $\checkmark$ & $\checkmark$ & $\times$     & $\times$     & $\times$ & $\times$ \\
\cite{shashidhar08}    & $\times$     & $\checkmark$ & $\times$     & $\times$     & $\times$ & $\times$ \\
\cite{yices,cvc4,acl2} & $\times$     & $\checkmark$ & $\times$     & Linear       & $\times$ & $\times$ \\
\cite{Verdoolaege2012} & $\times$     & $\checkmark$ & $\checkmark$ & $\times$     & $\times$ & $\times$ \\
\cite{karfaTcad13}     & $\times$     & $\checkmark$ & $\times$     & $\checkmark$ & $\times$ & $\times$ \\
\cite{lctes16}         & $\times$     & $\checkmark$ & $\checkmark$ & $\checkmark$ & $\times$ & $\times$ \\
\cite{array16}         & $\times$     & $\checkmark$ & $\checkmark$ & $\checkmark$ & $\times$ & $\checkmark$ \\
\cite{Dutta17}         & $\times$     & $\checkmark$ & $\checkmark$ & $\times$ & $\checkmark$ & $\checkmark$ \\
\hline           
\end{tabular}
\end{table*}

\input{refReport.bbl}

\end{document}

%% file: refReport.bbl

%% file: Banerjee_Karfa_Intro_Array_Program_Verification.bbl
\begin{thebibliography}{10}
\providecommand{\url}[1]{#1}
\csname url@samestyle\endcsname
\providecommand{\newblock}{\relax}
\providecommand{\bibinfo}[2]{#2}
\providecommand{\BIBentrySTDinterwordspacing}{\spaceskip=0pt\relax}
\providecommand{\BIBentryALTinterwordstretchfactor}{4}
\providecommand{\BIBentryALTinterwordspacing}{\spaceskip=\fontdimen2\font plus
\BIBentryALTinterwordstretchfactor\fontdimen3\font minus
  \fontdimen4\font\relax}
\providecommand{\BIBforeignlanguage}[2]{{%
\expandafter\ifx\csname l@#1\endcsname\relax
\typeout{** WARNING: IEEEtran.bst: No hyphenation pattern has been}%
\typeout{** loaded for the language `#1'. Using the pattern for}%
\typeout{** the default language instead.}%
\else
\language=\csname l@#1\endcsname
\fi
#2}}
\providecommand{\BIBdecl}{\relax}
\BIBdecl

\bibitem{openmp}
``{OpenMP},'' \url{https://www.openmp.org/}, [Online accessed: 5-Feb-2019].

\bibitem{openacc}
``{OpenACC},'' \url{https://www.openacc.org/}, [Online accessed: 5-Feb-2019].

\bibitem{mpi}
``{Message Passing Interface (MPI)},''
  \url{https://computing.llnl.gov/tutorials/mpi/}, [Online accessed:
  5-Feb-2019].

\bibitem{opencl}
``{OpenCL Overview},'' \url{https://www.khronos.org/opencl/}, [Online accessed:
  5-Feb-2019].

\bibitem{polaris}
W.~Blume, R.~Eigenmann, K.~Faigin, J.~Grout, J.~Hoeflinger, D.~A. Padua,
  P.~Petersen, W.~M. Pottenger, L.~Rauchwerger, P.~Tu, and S.~Weatherford,
  ``Polaris: Improving the effectiveness of parallelizing compilers,'' in
  \emph{LCPC}, 1994, pp. 141--154.

\bibitem{pluto}
U.~Bondhugula, A.~Hartono, J.~Ramanujam, and P.~Sadayappan, ``A practical
  automatic polyhedral parallelizer and locality optimizer,'' in \emph{PLDI},
  2008, pp. 101--113.

\bibitem{cgraNN1}
S.~M. A.~H. Jafri, T.~N. Gia, S.~Dytckov, M.~Daneshtalab, A.~Hemani,
  J.~Plosila, and H.~Tenhunen, ``Neurocgra: {A} {CGRA} with support for neural
  networks,'' in \emph{HPCS}, 2014, pp. 506--511.

\bibitem{cgraNN2}
M.~Tanomoto, S.~Takamaeda{-}Yamazaki, J.~Yao, and Y.~Nakashima, ``A
  {CGRA}-based approach for accelerating convolutional neural networks,'' in
  \emph{MCSoC}, 2015, pp. 73--80.

\bibitem{fpgaNN1}
Y.~Ma, Y.~Cao, S.~B.~K. Vrudhula, and J.~Seo, ``Optimizing the convolution
  operation to accelerate deep neural networks on {FPGA},'' \emph{{IEEE} Trans.
  {VLSI} Syst.}, vol.~26, no.~7, pp. 1354--1367, 2018.

\bibitem{systolicNN1}
\BIBentryALTinterwordspacing
A.~Samajdar, Y.~Zhu, P.~N. Whatmough, M.~Mattina, and T.~Krishna,
  ``{SCALE-Sim}: Systolic {CNN} accelerator,'' \emph{CoRR}, vol.
  abs/1811.02883, 2018. [Online]. Available:
  \url{http://arxiv.org/abs/1811.02883}
\BIBentrySTDinterwordspacing

\bibitem{xla}
``{XLA Overview},'' \url{https://www.tensorflow.org/xla/overview}, [Online
  accessed: 6-Feb-2019].

\bibitem{nnvm}
``{NNVM} compiler: Open compiler for {AI} frameworks,''
  \url{https://tvm.ai/2017/10/06/nnvm-compiler-announcement.html}, [Online
  accessed: 6-Feb-2019].

\bibitem{glow}
\BIBentryALTinterwordspacing
N.~Rotem, J.~Fix, S.~Abdulrasool, S.~Deng, R.~Dzhabarov, J.~Hegeman,
  R.~Levenstein, B.~Maher, N.~Satish, J.~Olesen, J.~Park, A.~Rakhov, and
  M.~Smelyanskiy, ``Glow: Graph lowering compiler techniques for neural
  networks,'' \emph{CoRR}, vol. abs/1805.00907, 2018. [Online]. Available:
  \url{http://arxiv.org/abs/1805.00907}
\BIBentrySTDinterwordspacing

\bibitem{T2S}
\BIBentryALTinterwordspacing
H.~Rong, ``Programmatic control of a compiler for generating high-performance
  spatial hardware,'' \emph{CoRR}, vol. abs/1711.07606, 2017. [Online].
  Available: \url{http://arxiv.org/abs/1711.07606}
\BIBentrySTDinterwordspacing

\bibitem{dlvm}
\BIBentryALTinterwordspacing
R.~Wei, V.~S. Adve, and L.~Schwartz, ``{DLVM:} {A} modern compiler
  infrastructure for deep learning systems,'' \emph{CoRR}, vol. abs/1711.03016,
  2017. [Online]. Available: \url{http://arxiv.org/abs/1711.03016}
\BIBentrySTDinterwordspacing

\bibitem{llvm}
``The llvm compiler infrastructure,'' \url{http://llvm.org/}.

\bibitem{337425}
M.~Kandemir, N.~Vijaykrishnan, M.~J. Irwin, and W.~Ye, ``Influence of compiler
  optimizations on system power,'' \emph{IEEE Trans. Very Large Scale Integr.
  Syst.}, vol.~9, pp. 801--804, 2001.

\bibitem{1077655}
M.~Kandemir, S.~W. Son, and G.~Chen, ``An evaluation of code and data
  optimizations in the context of disk power reduction,'' in \emph{ISLPED},
  2005, pp. 209--214.

\bibitem{1142163}
M.~T. Kandemir, ``Reducing energy consumption of multiprocessor {SoC}
  architectures by exploiting memory bank locality,'' \emph{ACM Trans. Des.
  Autom. Electron. Syst.}, vol.~11, no.~2, pp. 410--441, 2006.

\bibitem{Xue:2007:MCP:1278480.1278536}
L.~Xue, O.~Ozturk, and M.~Kandemir, ``A memory-conscious code parallelization
  scheme,'' in \emph{DAC}, 2007, pp. 230--233.

\bibitem{kadayifK05tecs}
I.~Kadayif and M.~T. Kandemir, ``Data space-oriented tiling for enhancing
  locality,'' \emph{ACM Trans. Embedded Comput. Syst.}, vol.~4, no.~2, pp.
  388--414, 2005.

\bibitem{li05workload}
F.~Li and M.~T. Kandemir, ``Locality-conscious workload assignment for
  array-based computations in {MPSOC} architectures,'' in \emph{DAC}, 2005, pp.
  95--100.

\bibitem{1118342}
G.~Chen, M.~Kandemir, and F.~Li, ``Energy-aware computation duplication for
  improving reliability in embedded chip multiprocessors,'' in \emph{ASP-DAC},
  2006, pp. 134--139.

\bibitem{bouchebaba07}
Y.~Bouchebaba, B.~Girodias, G.~Nicolescu, E.~M. Aboulhamid, B.~Lavigueur, and
  P.~G. Paulin, ``{MPSoC} memory optimization using program transformation,''
  \emph{ACM Trans. Design Autom. Electr. Syst.}, vol.~12, no.~4, 2007.

\bibitem{1241831}
C.~Zhang and F.~Kurdahi, ``Reducing off-chip memory access via stream-conscious
  tiling on multimedia applications,'' \emph{Int. J. Parallel Program.},
  vol.~35, no.~1, pp. 63--98, 2007.

\bibitem{memoryAlloc}
J.~Cong, P.~Zhang, and Y.~Zou, ``Combined loop transformation and hierarchy
  allocation for data reuse optimization,'' in \emph{ICCAD}, 2011, pp.
  185--192.

\bibitem{memoryAllocation}
------, ``Optimizing memory hierarchy allocation with loop transformations for
  high-level synthesis,'' in \emph{DAC}, 2012, pp. 1233--1238.

\bibitem{sanket11}
S.~Tavarageri, L.~Pouchet, J.~Ramanujam, A.~Rountev, and P.~Sadayappan,
  ``Dynamic selection of tile sizes,'' in \emph{HiPC}, 2011, pp. 1--10.

\bibitem{sanket13}
S.~Tavarageri, J.~Ramanujam, and P.~Sadayappan, ``Adaptive parallel tiled code
  generation and accelerated auto-tuning,'' \emph{{IJHPCA}}, vol.~27, no.~4,
  pp. 412--425, 2013.

\bibitem{1053675}
M.~Karakoy, ``Optimizing array-intensive applications for on-chip
  multiprocessors,'' \emph{IEEE Trans. Parallel Distrib. Syst.}, vol.~16,
  no.~5, pp. 396--411, 2005.

\bibitem{qiu08}
M.~Qiu, E.~H.-M. Sha, M.~Liu, M.~Lin, S.~Hua, and L.~T. Yang, ``Energy
  minimization with loop fusion and multi-functional-unit scheduling for
  multidimensional {DSP},'' \emph{J. Parallel Distrib. Comput.}, vol.~68,
  no.~4, pp. 443--455, 2008.

\bibitem{springerlink:10.1007}
M.~Ghodrat, T.~Givargis, and A.~Nicolau, ``Optimizing control flow in loops
  using interval and dependence analysis,'' \emph{Design Automation for
  Embedded Systems}, vol.~13, pp. 193--221, 2009.

\bibitem{Panda:2001:DMO:375977.375978}
P.~R. Panda, F.~Catthoor, N.~D. Dutt, K.~Danckaert, E.~Brockmeyer, C.~Kulkarni,
  A.~Vandercappelle, and P.~G. Kjeldsberg, ``Data and memory optimization
  techniques for embedded systems,'' \emph{ACM Trans. Des. Autom. Electron.
  Syst.}, vol.~6, pp. 149--206, April 2001.

\bibitem{baconSharp}
D.~F. Bacon, S.~L. Graham, and O.~J. Sharp, ``Compiler transformations for
  high-performance computing,'' \emph{ACM Comput. Surv.}, vol.~26, pp.
  345--420, 1994.

\bibitem{imecbk}
F.~Catthoor, E.~D., and S.~S. Greff, \emph{HICSS. Custom Memory Management
  Methodology: Exploration of Memory Organisation for Embedded Multimedia
  System Design}.\hskip 1em plus 0.5em minus 0.4em\relax Kluwer Academic
  Publishers, 1998.

\bibitem{palkovic09}
M.~Palkovic, F.~Catthoor, and H.~Corporaal, ``Trade-offs in loop
  transformations,'' \emph{ACM Trans. Design Autom. Electr. Syst.}, vol.~14,
  no.~2, 2009.

\bibitem{FrabouletKM01}
A.~Fraboulet, K.~Kodary, and A.~Mignotte, ``Loop fusion for memory space
  optimization,'' in \emph{ISSS}, 2001, pp. 95--100.

\bibitem{loopCgra}
D.~Liu, S.~Yin, L.~Liu, and S.~Wei, ``Polyhedral model based mapping
  optimization of loop nests for cgras,'' in \emph{DAC}, 2013, pp. 19:1--19:8.

\bibitem{7084524}
S.~Prema, R.~Jehadeesan, B.~Panigrahi, and S.~Satya~Murty, ``Dependency
  analysis and loop transformation characteristics of auto-parallelizers,'' in
  \emph{Parallel Computing Technologies}, 2015, pp. 1--6.

\bibitem{loopMultiCore}
H.~Yviquel, A.~Sanchez, P.~J{\"{a}}{\"{a}}skel{\"{a}}inen, J.~Takala,
  M.~Raulet, and E.~Casseau, ``Embedded multi-core systems dedicated to dynamic
  dataflow programs,'' \emph{Signal Processing Systems}, vol.~80, no.~1, pp.
  121--136, 2015.

\bibitem{Simunic:2000:SCO:501790.501831}
T.~\v{S}imuni\'{c}, L.~Benini, G.~De~Micheli, and M.~Hans, ``Source code
  optimization and profiling of energy consumption in embedded systems,'' in
  \emph{International Symposium on Systems Synthesis}, 2000, pp. 193--198.

\bibitem{1026005}
Y.~Zhu, G.~Magklis, M.~L. Scott, C.~Ding, and D.~H. Albonesi, ``The energy
  impact of aggressive loop fusion,'' in \emph{PACT}, 2004, pp. 153--164.

\bibitem{Brandolese:2004:AME:968880.969253}
C.~Brandolese, W.~Fornaciari, F.~Salice, and D.~Sciuto, ``Analysis and modeling
  of energy reducing source code transformations,'' in \emph{DATE}, 2004, pp.
  306--311.

\bibitem{Ghodrat:2008:CFO:1450095.1450120}
M.~Ghodrat, T.~Givargis, and A.~Nicolau, ``Control flow optimization in loops
  using interval analysis,'' in \emph{Proceedings of the 2008 international
  conference on Compilers, architectures and synthesis for embedded systems},
  2008, pp. 157--166.

\bibitem{falkbook}
H.~Falk and P.~Marwedel, \emph{Source code optimization techniques for data
  flow dominated embedded software}.\hskip 1em plus 0.5em minus 0.4em\relax
  Kluwer, 2004.

\bibitem{potkonjakRetiming}
M.~Potkonjak, S.~Dey, Z.~Iqbal, and A.~C. Parker, ``High performance embedded
  system optimization using algebraic and generalized retiming techniques,'' in
  \emph{ICCD}, 1993, pp. 498--504.

\bibitem{zoryPACT}
J.~Zory and F.~Coelho, ``Using algebraic transformations to optimize expression
  evaluation in scientific codes,'' in \emph{PACT}, 1998, pp. 376--384.

\bibitem{sguptaDATE00}
S.~Gupta, R.~K. Gupta, M.~Miranda, and F.~Catthoor, ``Analysis of high-level
  address code transformations for programmable processors,'' in \emph{DATE},
  2000, pp. 9--13.

\bibitem{landwehr97}
B.~Landwehr and P.~Marwedel, ``A new optimization technique for improving
  resource exploitation and critical path minimization,'' in \emph{ISSS}, 1997,
  pp. 65--72.

\bibitem{spark03}
S.~Gupta, N.~Dutt, R.~Gupta, and A.~Nicolau, ``{SPARK}: A high-level synthesis
  framework for applying parallelizing compiler transformations,'' in
  \emph{VLSI Design}, 2003, pp. 461--466.

\bibitem{sGuptaACM}
S.~Gupta, R.~Gupta, N.~Dutt, and A.~Nicolau, ``Coordinated parallelizing
  compiler optimizations and high-level synthesis,'' \emph{ACM Trans. on Design
  Autom. of Electr. Syst.}, vol.~9, no.~4, pp. 1--31, October 2004.

\bibitem{eoapndg03}
A.~P.~N. E.~\"{O}zer and D.~Gregg, ``Classification of compiler optimizations
  for high performance, small area and low power in fpgas,'' Trinity College,
  Dublin, Ireland, Department of Computer Science, Tech. Rep., 2003.

\bibitem{voc}
L.~Zuck, A.~Pnueli, Y.~Fang, and B.~Goldberg, ``{VOC}: A translation validator
  for optimizing compilers,'' \emph{Journal of Universal Computer Science},
  vol.~9, no.~3, pp. 223--247, 2003.

\bibitem{pnueli05}
L.~D. Zuck, A.~Pnueli, B.~Goldberg, C.~W. Barrett, Y.~Fang, and Y.~Hu,
  ``Translation and run-time validation of loop transformations,'' \emph{Formal
  Methods in System Design}, vol.~27, no.~3, pp. 335--360, 2005.

\bibitem{loopEntcs}
Y.~Hu, C.~W. Barrett, B.~Goldberg, and A.~Pnueli, ``Validating more loop
  optimizations,'' \emph{Electr. Notes Theor. Comput. Sci.}, vol. 141, no.~2,
  pp. 69--84, 2005.

\bibitem{tvoc}
C.~W. Barrett, Y.~Fang, B.~Goldberg, Y.~Hu, A.~Pnueli, and L.~D. Zuck,
  ``{TVOC}: A translation validator for optimizing compilers,'' in \emph{CAV},
  2005, pp. 291--295.

\bibitem{fractal}
V.~Menon, K.~Pingali, and N.~Mateev, ``Fractal symbolic analysis,'' \emph{ACM
  Trans. Program. Lang. Syst.}, vol.~25, no.~6, pp. 776--813, 2003.

\bibitem{samsom}
H.~Samsom, F.~Franssen, F.~Catthoor, and H.~De~Man, ``System level verification
  of video and image processing specifications,'' in \emph{International
  Symposium on Systems Synthesis}, 1995, pp. 144--149.

\bibitem{cvc4}
``{CVC4 - the smt solver},'' \url{http://cvc4.cs.nyu.edu/web/}.

\bibitem{yices}
``{The Yices SMT Solver},'' \url{http://yices.csl.sri.com/}.

\bibitem{acl2}
``{ACL2 Version 6.1},'' \url{http://www.cs.utexas.edu/~moore/acl2/}.

\bibitem{icare13}
C.~Karfa, K.~Banerjee, D.~Sarkar, and C.~Mandal, ``Experimentation with {SMT}
  solvers and theorem provers for verification of loop and arithmetic
  transformations,'' in \emph{I-CARE}, 2013, pp. 3:1--3:4.

\bibitem{kcs02}
K.~C. Shashidhar, M.~Bruynooghe, F.~Catthoor, and G.~Janssens, ``Geometric
  model checking: An automatic verification technique for loop and data reuse
  transformations,'' \emph{Electronic Notes in Theoretical Computer Science},
  vol.~65, no.~2, pp. 71--86, 2002.

\bibitem{kcscc05}
------, ``Verification of source code transformations by program equivalence
  checking,'' in \emph{Compiler Construction}, 2005, pp. 221--236.

\bibitem{shashidhar05}
------, ``Functional equivalence checking for verification of algebraic
  transformations on array-intensive source code,'' in \emph{DATE}, 2005, pp.
  1310--1315.

\bibitem{shashidhar08}
K.~C. Shashidhar, ``Efficient automatic verification of loop and data-flow
  transformations by functional equivalence checking,'' Ph.D. dissertation,
  Katholieke Universiteit Leuven, 2008.

\bibitem{wideningCAV09}
S.~Verdoolaege, G.~Janssens, and M.~Bruynooghe, ``Equivalence checking of
  static affine programs using widening to handle recurrences,'' in \emph{CAV},
  2009, pp. 599--613.

\bibitem{widening12}
------, ``Equivalence checking of static affine programs using widening to
  handle recurrences,'' \emph{ACM Trans. Program. Lang. Syst.}, vol.~34, no.~3,
  2012.

\bibitem{isvlsi11}
C.~Karfa, K.~Banerjee, D.~Sarkar, and C.~Mandal, ``Equivalence checking of
  array-intensive programs,'' in \emph{ISVLSI}, 2011, pp. 156--161.

\bibitem{karfaTcad13}
------, ``Verification of loop and arithmetic transformations of
  array-intensive behaviours,'' \emph{IEEE Trans. on CAD of ICS}, vol.~32,
  no.~11, pp. 1787--1800, 2013.

\bibitem{widening_J10}
S.~Verdoolaege, M.~Palkovic, M.~Bruynooghe, G.~Janssens, and F.~Catthoor,
  ``Experience with widening based equivalence checking in realistic multimedia
  systems,'' \emph{J. Electronic Testing}, vol.~26, no.~2, pp. 279--292, 2010.

\bibitem{dsNorm}
D.~Sarkar and S.~{De Sarkar}, ``A theorem prover for verifying iterative
  programs over integers,'' \emph{IEEE Trans Software. Engg.}, vol.~15, no.~12,
  pp. 1550--1566, 1989.

\bibitem{karfaVlsid13}
C.~Karfa, D.~Sarkar, and C.~A. Mandal, ``Verification of {KPN} level
  transformations,'' in \emph{VLSI Design}, 2013, pp. 338--343.

\bibitem{lctes16}
K.~Banerjee, C.~Mandal, and D.~Sarkar, ``Translation validation of loop and
  arithmetic transformations in the presence of recurrences,'' in \emph{LCTES},
  2016, pp. 31--40.

\bibitem{Flanagan2002}
C.~Flanagan, S.~N. Freund, and S.~Qadeer, ``{Thread-Modular Verification for
  Shared-Memory Programs},'' \emph{Programming Languages and Systems, LNCS},
  vol. 2305, pp. 262--277, 2002.

\bibitem{Deng2002}
X.~Deng, M.~B. Dwyer, J.~Hatcliff, and M.~Mizuno, ``{Invariant-based
  Specification, Synthesis, and Verification of Synchronization in Concurrent
  Programs},'' in \emph{ICSE}, 2002, pp. 442--452.

\bibitem{array16}
K.~Banerjee, S.~Banerjee, and S.~Sarkar, ``Data-race detection: the missing
  piece for an end-to-end semantic equivalence checker for parallelizing
  transformations of array-intensive programs,'' in \emph{ARRAY@PLDI}, 2016,
  pp. 1--8.

\bibitem{fase15}
S.~Blom, S.~Darabi, and M.~Huisman, ``Verification of loop parallelisations,''
  in \emph{FASE}, 2015, pp. 202--217.

\bibitem{dutta16}
S.~Dutta, D.~Sarkar, A.~Rawat, and K.~Singh, ``Validation of loop
  parallelization and loop vectorization transformations,'' in \emph{{ENASE}},
  2016, pp. 195--202.

\bibitem{Dutta17}
S.~Dutta, ``Validation of parallelizing transformations of sequential
  programs,'' \emph{Concurrency and Computation: Practice and Experience},
  vol.~29, no.~8, 2017.

\bibitem{Verdoolaege2012}
S.~Verdoolaege, G.~Janssens, and M.~Bruynooghe, ``Equivalence checking of
  static affine programs using widening to handle recurrences,'' \emph{{ACM}
  TOPLAS}, vol.~34, no.~3, p.~11, 2012.

\end{thebibliography}
